# The Triple Helix of University-Industry-Government Relations at the Country Level, and its Dynamic Evolution under the Pressures of Globalization

Fred Y. Ye,[a*]   Susan S. Yu, [b] & Loet Leydesdorff [c]

[a]School of Information Management, Nanjing University, Nanjing, China; *yye@nju.edu.cn;

[b]Dept. of Information Resource Management, Zhejiang University, Hangzhou, China;

[c]Amsterdam School of Communication Research (ASCoR), University of Amsterdam,

Kloveniersburgwal 48, 1012 CX Amsterdam, The Netherlands; loet@leydesdorff.net

**Abstract:** Using data from the Web of Science (WoS), we analyze the mutual information among university, industrial, and governmental addresses (U-I-G) at the country level for a number of countries. The dynamic evolution of the Triple Helix can thus be compared among developed and developing nations in terms of cross-sectorial co-authorship relations. The results show that the Triple-Helix interactions among the three subsystems U-I-G become less intensive over time, but unequally for different countries. We suggest that globalization erodes local Triple-Helix relations and thus can be expected to increase differentiation in national systems since the mid-1990s. This effect of globalization is more pronounced in developed countries than in developing ones. In the dynamic analysis, we focus on a more detailed comparison between China and the USA. The Chinese Academy of the (Social) Sciences changes increasingly from a public research institute to an academic one, and this has a measurable effect on China's position in the globalization.

**Introduction**

The Triple Helix model of university-industry-government relations was introduced in 1995 by Etzkowitz & Leydesdorff (Etzkowitz & Leydesdorff, 1995, 2000; Leydesdorff & Etzkowitz, 1996 and 1998). This model has been widely used since then, particularly in studies of the knowledge-based economy and innovation (e.g., Etzkowitz, 2008; Jacob, 2006; Leydesdorff, 2006; 2010; Mirowski & Sent, 2010). Leydesdorff (2003: 458) proposed a scientometric operationalization of the Triple Helix thesis in terms of the mutual information in three dimensions—alternatively named, configurational information. Park *et al.* (2005) used this Triple-Helix indicator for comparing the progression of the knowledge-based economy in South Korea with other countries.

In this study, we follow previous studies about Korea (Kwon *et al.*, 2012; Park & Leydesdorff, 2010; Park *et al.*, 2005) and Japan (Leydesdorff & Sun, 2009), and analyze triple-helix relations and their dynamic evolution at the country level using Web of Science (WoS) address information. Institutional coauthorship relations can be retrieved from this database as a proxy of collaborations (e.g., Glänzel & Schubert, 2005; Zitt *et al.*, 2000; Wagner & Leydesdorff, 2005). Although our search strings are pragmatic, the macro-results show patterns in the development of these relations that can be considered as the effects of globalization: national exchange relations among sectorially different institutions become gradually less important than functional relations within sectors. Functional relations, however, can be expected to extend beyond national borders (Kwon *et al.*, 2012).

More than other Asian nations, China follows this "western" pattern of differentiation in terms of (co)authorship relations. In other words, its Triple-Helix relationships have become less nationally oriented. We also raise the question of whether this result is sensitive to the attribution of publications of the Chinese Academy of the Sciences (CAS) and the Chinese Academy of the Social Sciences (CASS) to the university or the governmental system of publications. Since CAS and CASS are funded as governmental agencies, they are often considered as public research institutes (i.e., "CAS as G"). However, CAS has in the meantime also organized a graduate school (and accordingly changed its name into the University of the Chinese Academy), which belongs to university (i.e., "CAS as U"). We distinguish between the two attributions, and discuss their possible effect on UIG relations.

Meanwhile, we also mention the wide applications of Trip Helix theory, particularly in innovation studies (Etzkowitz, 2003) and funding management (Benner & Sandström, 2000), as well as knowledge production in science and technology (Shinn, 2002).

**Methodology**

*a. The TH-indicator as a **signed** information measure*

Let us first clarify the logic of relations among U, I and G, and mutual or configurational information. One can distinguish between two Venn charts as representations of the data using the logical operators OR or AND, respectively (Figures 1 and 2).

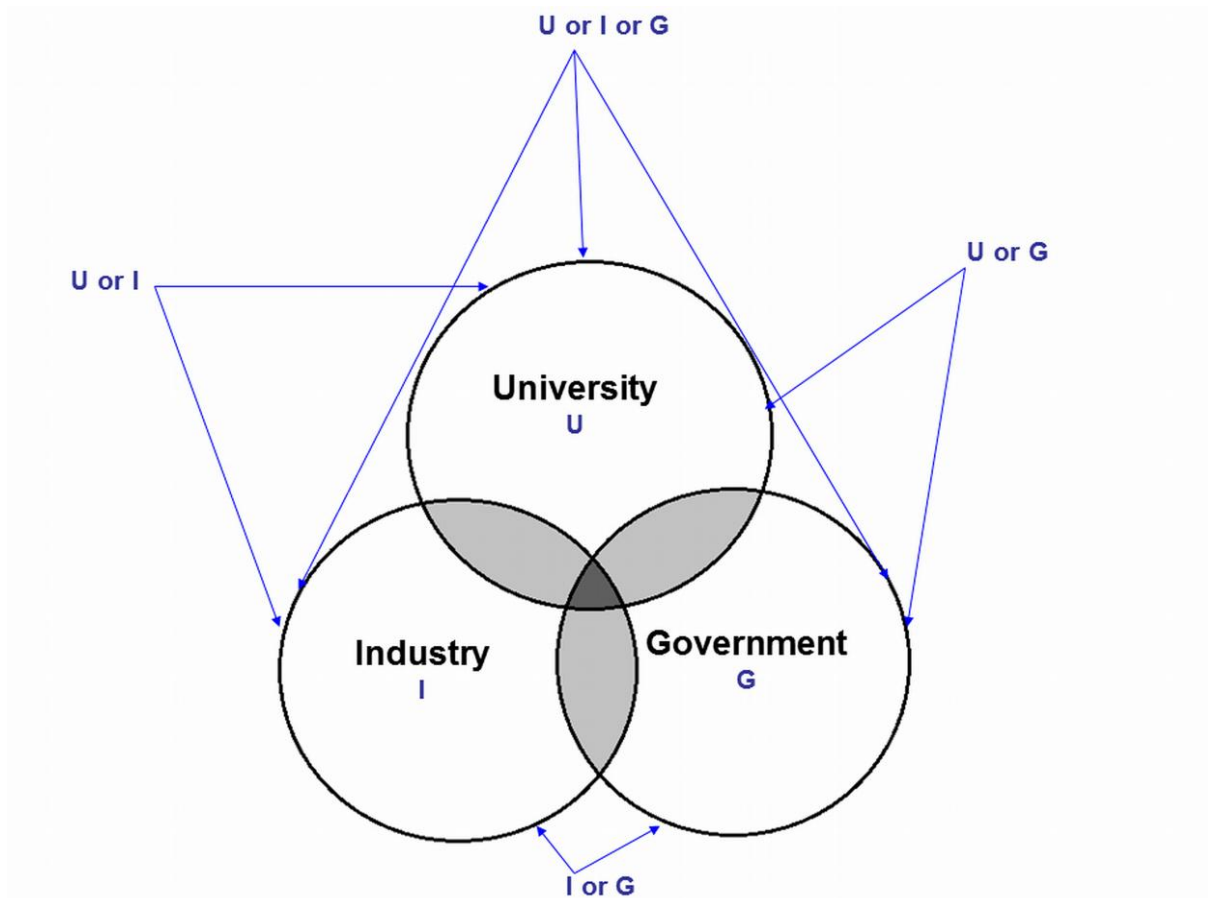

**Fig. 1**. Joint entropy in a triple helix system using the logical operator OR

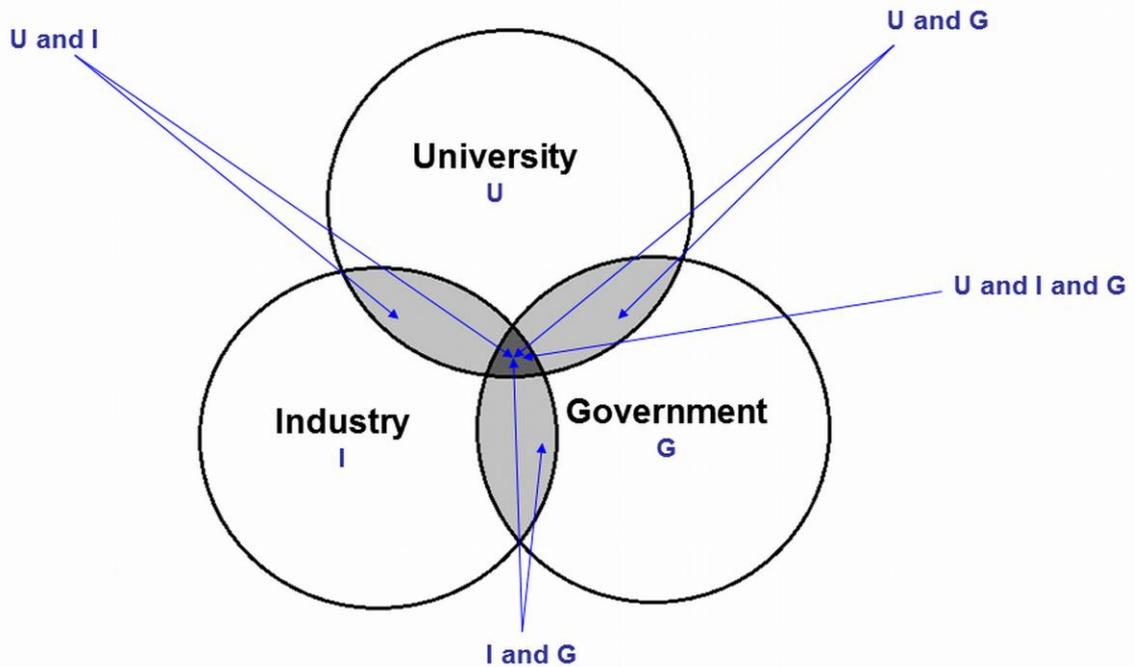

**Fig. 2**. Mutual information in a triple helix system using the logical operator AND

The retrieval provides us with sets containing relative frequency distributions in these different domains. Each relative frequency distribution contains an uncertainty that can be measured using Shannon's (1948) formulas in terms of bits of information, as follows:

$$H(X) = -\sum_{x \in X} p(x) \log p(x) \qquad (1)$$

$H(X)$ is the uncertainty in—or, in other words, the probabilistic entropy of—a discrete random variable $X$. The value of Shannon's $H$ is always positive since $1 \geq p(x) \geq 0$ for all $p(x)$. When two is used as the basis for the logarithm, uncertainty is expressed in bits. (One can freely change the base of the logarithm by using the formula: $\log_a p = \log_b p / \log_a p$.)

Joint entropy, conditional entropy, and mutual information are straightforward extensions of $H$ that measure uncertainty in the joint distribution of a pair of random variables X and Y. The difference between the uncertainty in the distribution X and the conditional distribution X|Y is equal to the transmission or mutual information between X

and Y:

$$T(X) = H(X) - H(X/Y) \qquad (2)$$

The joint entropy $H(X,Y)$ of a pair of discrete random variables with a joint distribution $p(x, y)$ is defined as

$$H(X,Y) = -\sum_{x \in X} \sum_{y \in Y} p(x, y) \log p(x, y) \qquad (3)$$

The relation between joint entropy and conditional entropy is given by the so-called chain rule theorem that is formalized as follows:

$$H(X, Y) = H(X) + H(Y/X) = H(Y) + H(X/Y) \qquad (4)$$

The relations are symmetrical: the mutual information or transmission $T(X, Y)$ can from this perspective also be considered as follows:

$$T(X, Y) = H(X) + H(Y) - H(X, Y) \qquad (5)$$

These various relations are summarized in Figure 3.

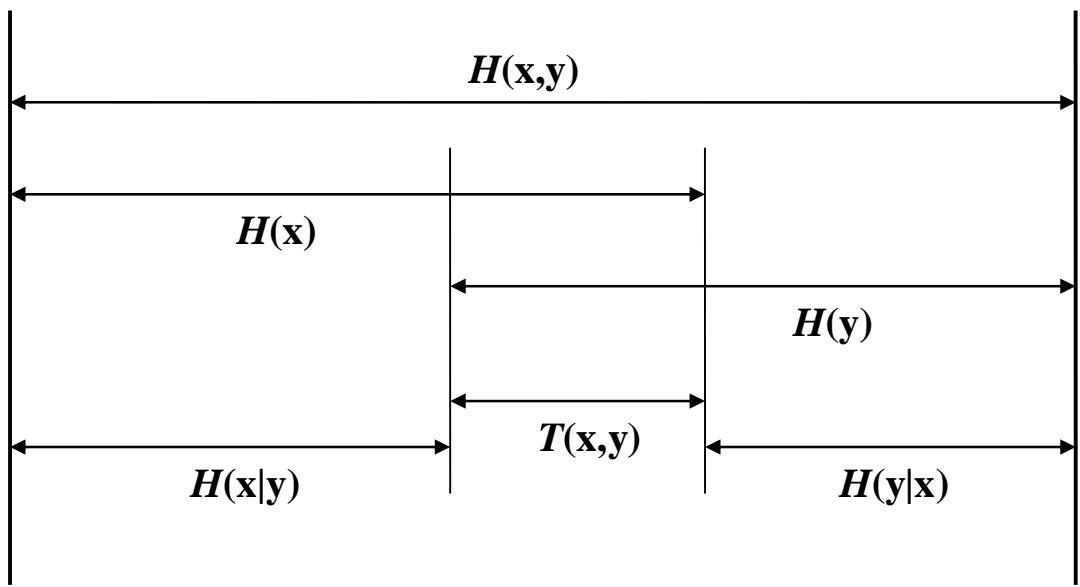

**Fig. 3**: Relations of expected information contents, mutual information, and conditional entropies between two variables *x* and *y*. (Source: Attneave, 1959, p. 49.)

Furthermore, one can derive that the mutual information in three dimensions—U, I, and G in our study—is defined as follows (McGill, 1954; Yeung, 2009: 59f.):

$$T(UIG) = H(U) + H(I) + H(G) - H(UI) - H(UG) - H(IG) + H(UIG) \qquad (6)$$

where *H(XY)* means *H(X,Y)*. A negative value of *T(UIG)* indicates a reduction of the uncertainty that prevails and can therefore be considered as a measure of the *synergy* in UIG relations.

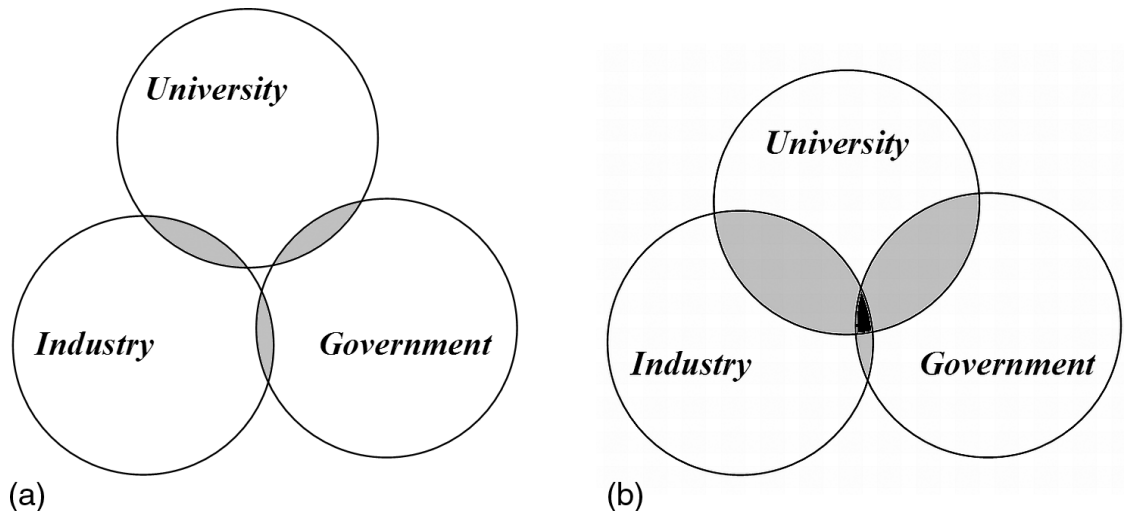

**Fig. 4**. Negative and positive overlaps among U-I-G relations.

We are interested in the relations between negative and positive overlap among the three subsystems U-I-G, as alternating possibilities (Figure 4). Using the Triple-Helix indicator as a *signed* measure, we are able to measure the Triple Helix configurations and their dynamic evolution at the country level in terms of coauthorship relations at the institutional level. More negative values indicate a reduction of uncertainty or, in other words, a synergy, whereas more positive values indicate differentiation among the three spheres.

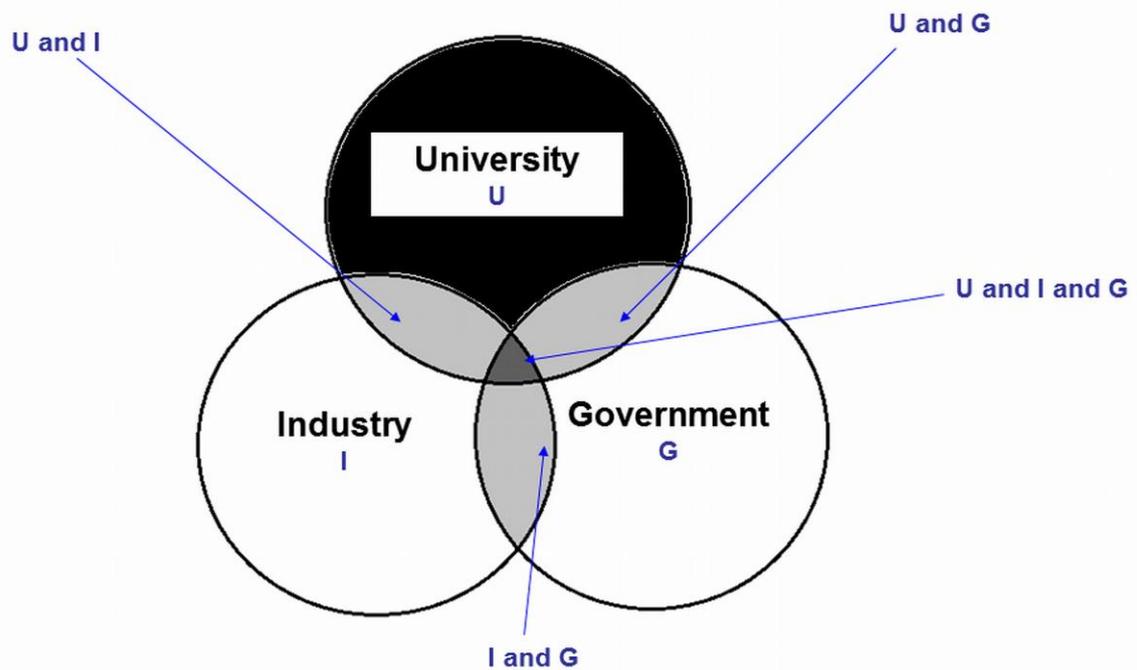

**Fig. 5**. Configurational information in a Triple Helix system

Let us explain this reduction of uncertainty by a third system in more detail using Figure 5. The relation between I and G at the bottom of Figure 5 may be partially and/or spuriously correlated by a third system U. In the case of partial correlation, the mutual information in three dimensions is positive, and in the case of a spurious correlation, the third factor operates as a latent variable that can reduce uncertainty in the *system of relations* (Strand & Leydesdorff, in press). Such a reduction of uncertainty is indicated by a negative value. Krippendorff (2009a and b) showed that this redundancy (= negative information) can no longer be considered as a Shannon-type information because of the negative sign. Yeung (2009: 59f.) therefore called it a *signed* information measure.

*b. The retrieval*

Although Figures 2 and 5 may seem similar, they are different because each overlapping area is more than once retrieved (in Figure 2), while it should be counted only once (in Figure 5). Thus, we have first to correct the relative frequency distributions retrieved for double-counting of the overlaps.

Let $U_0$, $I_0$, $G_0$, $UI_0$, $UG_0$, $IG_0$ and $UIG_0$ be counts in the retrieval according to Fig. 2, and U, I, G, UI, UG, IG and UIG be counts to be used as the sources of configurational information (Fig. 5), then the following set of formulas provide the relations between the former and latter counts:

$$U = U_0 - UI_0 - UG_0 + UIG_0 \qquad (6)$$

$$I = I_0 - UI_0 - IG_0 + UIG_0 \qquad (7)$$

$$G = G_0 - IG_0 - UG_0 + UIG_0 \qquad (8)$$

$$UI = UI_0 - UIG_0 \qquad (9)$$

$$IG = IG_0 - UIG_0 \qquad (10)$$

$$UG = UG_0 - UIG_0 \qquad (11)$$

$$UIG = UIG_0 \qquad (12)$$

*c. Data*

Based on the logic of Figure 2 and using search strings previously developed and tested by Leydesdorff (2003: 458) and (Park *et al.*, 2005: 13 ff.), our search strategies were formulated as follows:

(1) $U_0$:

PY=year-year AND AD=(COUNTRY SAME (UNIV* OR COLL*))

This search string combines the addresses of universities and colleges with country names.

(2) $I_0$:

PY=year-year AND AD=(COUNTRY SAME (GMBH* OR CORP* OR LTD* OR AG*))

This search string combines country names with standard abbreviations for company addresses in English and German.

Thereafter, one needs the following intermediate steps:

(3) Intermediate step 1:

PY=year-year AND AD=(COUNTRY SAME (NATL* OR NACL* OR NAZL* OR GOVT* OR MINIST* OR ACAD* OR NIH*))

The search string combines country names with abbreviations for governmental and public research units

(4) Intermediate step 2:

PY=year-year AND AD=(COUNTRY SAME (NATL* OR NACL* OR NAZL* OR GOVT* OR MINIST* OR ACAD* OR NIH*) SAME (UNIV* OR COLL*))

(5) Intermediate step 3:

PY=year-year AND AD=(COUNTRY SAME (NATL* OR NACL* OR NAZL* OR GOVT* OR MINIST* OR ACAD* OR NIH*) SAME (GMBH* OR CORP* OR LTD* OR AG*))

In order to derive:

(6) $G_0$:    #3 NOT #4 NOT #5

(7) $UI_0$:    #1 AND #2

(8) $UG_0$:    #1 AND #6

(9) $IG_0$:    #2 AND #6

(10) $UIG_0$:    #1 AND #2 AND #6

The search strings 1-5 can be further enriched by adding more abbreviations with OR-statements between the brackets, but we did not pursue this further refinement in this study (Doranov & Leydesdorff, in preparation).

Data was collected for the various countries which belong to the G7 (Canada, France, Germany, Italy, Japan, UK, USA), BRICS (Brazil, Russia, India, China, South Africa), and a group which we define as INS (Indonesia, Netherlands, South Korea). South-Korea and the Netherlands were compared by Park *et al.* (2005) in a previous study and Indonesia was included because this country hosted the Triple Helix conference in 2012 (Durrani *et al.*, 2012).

Data was collected first for all these countries with publication year 2011 as the last available full year at the time of this research. Time series were collected for the period 1971-2010 with five-year time windows (e.g, PY = 2001-2005).

**Results**

*a.   The static comparison for 2011*

Using the above Equations 1-6, one can compute configurational information for the various countries as shown in the right column of Table 1.

Table 1. Shannon-type and configurational information for G7 + BRICS + INS in mbits of information, in 2011

Table 1. Shannon-type and configurational information for G7 + BRICS + INS in mbits of information, in 2011

| Country | H(U) | H(I) | H(G) | H(UI) | H(IG) | H(UG) | H(UIG) | T(UIG) |
|---|---|---|---|---|---|---|---|---|
| USA | 254.1 | 215.4 | 362.9 | 451 | 578.1 | 508.6 | 675.4 | **-29.96** |
| UK* | 225.3 | 280.3 | 186.9 | 439.1 | 467 | 362.3 | 542.5 | **-33.4** |
| FRANCE | 359.6 | 299.7 | 354.9 | 581.5 | 654.2 | 611.3 | 769.5 | **-63.17** |
| GERMANY | 306.7 | 468.8 | 52.36 | 598.1 | 521.1 | 351.4 | 626.5 | **-16.09** |
| ITALY | 300.6 | 285.5 | 476 | 575.1 | 759.1 | 655.7 | 894.8 | **-32.93** |
| CANADA | 213.4 | 321.1 | 188.4 | 474.1 | 509.6 | 361.1 | 591.7 | **-30.1** |
| JAPAN | 473 | 710.5 | 548.4 | 1138 | 1258 | 940.2 | 1529 | **-75.4** |
| BRAZIL | 206.9 | 455.9 | 298.3 | 647.3 | 752.1 | 444.3 | 855.4 | **-27.11** |
| RUSSIA | 998.1 | 157 | 972.9 | 1153 | 1120 | 1508 | 1595 | **-58.16** |
| INDIA | 650.9 | 480.7 | 654.5 | 1099 | 1128 | 1027 | 1359 | **-109.5** |
| CHINA(CAS as G) | 410.5 | 397.1 | 604.8 | 805.3 | 999 | 830.8 | 1187 | **-36.01** |
| CHINA(CAS as U) | 152.2 | 397.1 | 252.5 | 537 | 649.2 | 359.8 | 724.9 | **-19.45** |
| SOUTH AFRICA | 176.1 | 334.5 | 338.4 | 495.7 | 672.1 | 464.4 | 761.9 | **-21.35** |
| INDONESIA | 492.9 | 704.4 | 342.4 | 1120 | 1042 | 730.7 | 1270 | **-83.17** |
| NERHERLANDS | 172.8 | 217.9 | 242.1 | 365.7 | 459.3 | 358.3 | 527.6 | **-22.92** |
| SOUTH KOREA | 179.8 | 359.6 | 190.1 | 519.2 | 549.3 | 304.5 | 620.8 | **-22.55** |

*In WoS, UK=(England OR Scotland OR Wales OR North Ireland).

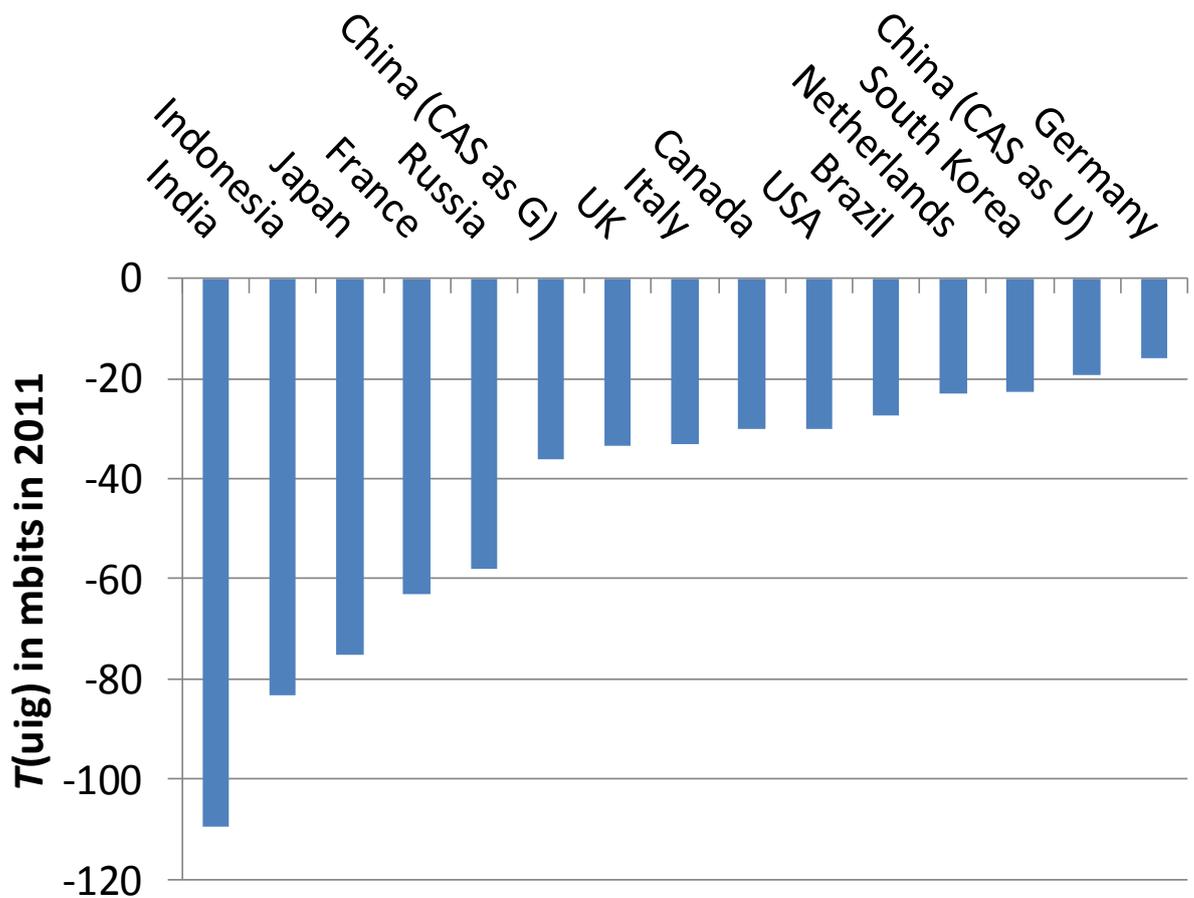

Fig. 6. T(UIG) distribution for the various countries in 2011.

In Figure 6 the values of T(UIG) are ranked in terms of decreasing negative values. At the left side of the figure, we find countries with a nationally oriented publication system, while at the right end the synergy in terms of Triple Helix relations—here expressed as a negative uncertainty—is much lower. These latter countries are more internationally oriented. Among them are small countries such as Korea and the Netherlands, but the most globally oriented country using this indicator is Germany. Interestingly, China is ranked second among these countries in terms of globalization if CAS(S) publications are counted as university publications, but China is ranked between Russia and the UK if CAS(S) publications are counted as output of governmental agencies. China inherited the Academy system from the Soviet Union.

The large Asian nations such as India, Indonesia, and Japan are on this scale the most inward-oriented nations in terms of Triple-Helix relations among institutions in different spheres. Perhaps, in these countries the policy emphasis should be more on differentiation and globalization than on further strengthening national integration. We shall see in the next section that China moved since the 1980s from a position similar to Indonesia toward one more close to South-Korea and the advanced industrial nations.

*b. The dynamic analysis using the time-series 1971-2010*

When the data is collected in five-year time steps for the entire period 1971-2010, we obtain the lists provided in Table 2. Note that before 1991, Germany was identified in this database as the Federal Republic of Germany—excluding the German Democratic Republic—and Russia was listed as the Soviet Union.

**Table 2**. Development of the configurational information of the G7, BRICS, and INS countries, during the period 1971-2010 (mbits of information)

| | USA | UK | FRANCE | GERMANY | ITALY | CANADA |
|---|---|---|---|---|---|---|
| 1971-1975 | -82.03 | -104.2 | -120.9 | -25.41 | -29.75 | -105.1 |
| 1976-1980 | -88.34 | -101.8 | -115.6 | -93.11 | -24.83 | -111.8 |
| 1981-1985 | -89.75 | -96.07 | -128.2 | -40.21 | -25.97 | -106.3 |
| 1986-1990 | -85.7 | -81.15 | -119.1 | -109.3 | -29.37 | -87.51 |
| 1991-1995 | -92.37 | -77.05 | -105.4 | -28.33 | -29.59 | -68.27 |
| 1996-2000 | -53.03 | -47.69 | -98.43 | -22.71 | -28.89 | -49.81 |
| 2001-2005 | -43.18 | -39.77 | -93.58 | -22.55 | -32.05 | -43.65 |
| 2006-2010 | -33.71 | -34.41 | -72.72 | -18.35 | -31.71 | -35.22 |
| | JAPAN | BRAZIL | RUSSIA | INDIA | CHINA (CAS as G) | CHINA (CAS as U) |
| 1971-1975 | -113.2 | -52.05 | *n.a.* | -101.9 | -173.9 | -173.9 |
| 1976-1980 | -114.8 | -68.52 | *n.a.* | -95.23 | -76.78 | -79.16 |
| 1981-1985 | -116.5 | -118.4 | *n.a.* | -106.7 | -80.53 | -80.11 |
| 1986-1990 | -114.3 | -106.3 | *n.a.* | -113.1 | -55.75 | -49.68 |
| 1991-1995 | -106.6 | -79.09 | -61.54 | -118.7 | -47.13 | -44.35 |
| 1996-2000 | -96.55 | -52.32 | -54.46 | -125.3 | -40.03 | -28.83 |
| 2001-2005 | -87.16 | -38.27 | -45.23 | -124.2 | -30.29 | -15.34 |
| 2006-2010 | -80.01 | -30.97 | -56.92 | -118.7 | -32.11 | -15.87 |
| | SOUTH AFRICA | INDO-NESIA | NETHER-LANDS | SOUTH KOREA | | |
| 1971-1975 | -79.44 | -229.7 | -26.16 | -252.6 | | |
| 1976-1980 | -97.29 | -157.9 | -38.24 | -48.52 | | |
| 1981-1985 | -88.21 | -157.6 | -44.66 | -45.96 | | |
| 1986-1990 | -69.52 | -233 | -43.38 | -17.21 | | |
| 1991-1995 | -47.32 | -135.9 | -53.3 | -27.9 | | |
| 1996-2000 | -36.88 | -132.7 | -47.1 | -19.91 | | |
| 2001-2005 | -32.17 | -125 | -35.58 | -23.81 | | |
| 2006-2010 | -24.82 | -115 | -29.76 | -25.91 | | |

Table 2 shows that most countries tend towards lower values (in absolute terms) of the Triple-Helix indicator over time, but not all. In India, for example, this value has not become less negative, but in South-Africa and Brazil such a tendency has taken hold. Some Western-European nations (Italy, Germany, and the Netherlands) have always had a low value on this indicator because of cross-border relations in the European context. This is less the case for leading nations such as the UK and France, but the values for these two nations have also been moderated during the 1990s and 2000s.

### c. Comparison of the developments in China and the USA

Let us, as an example, contrast the USA and China in more detail. This comparison allows us to focus on the different dynamics in developed and developing countries. Figures 7 and 8 teach us first that in the case of the USA, the dynamics changed profoundly when the period 1996-2010 is compared with 1971-1995. University publishing rapidly expanded since that time, and local integration in terms of university-industry-government relations became much less synergetic.

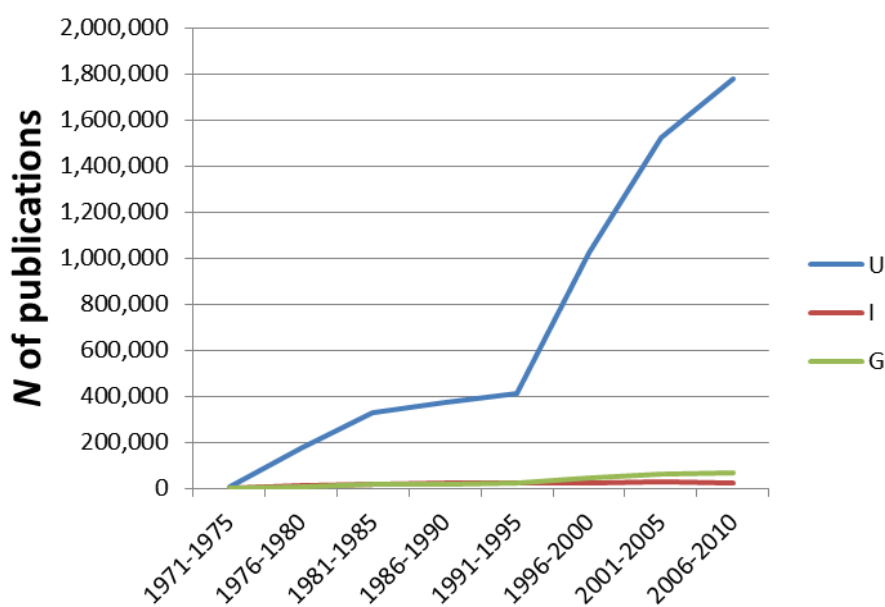

**Fig. 7**. Publication output of the USA in terms of U, I, and G; 1971-2010.

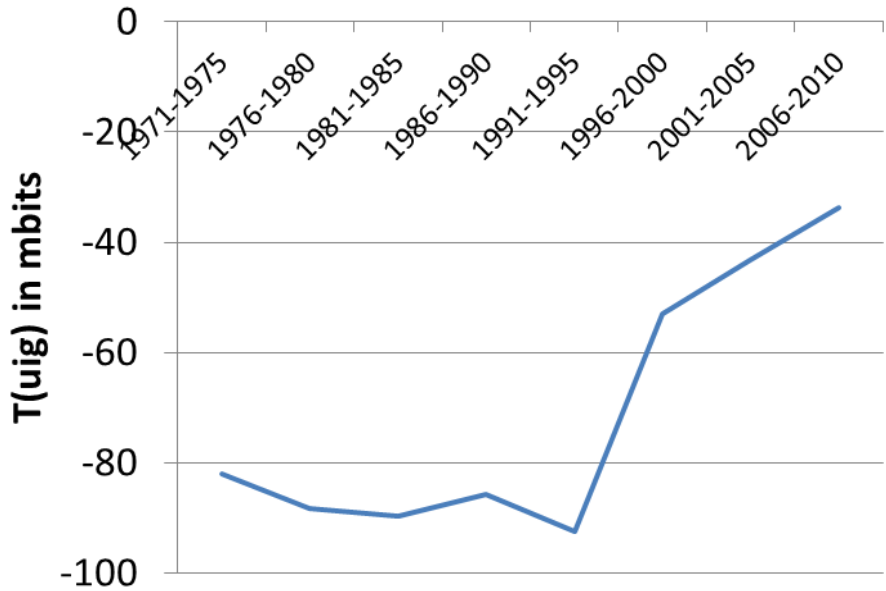

**Fig. 8**. The development of configurational information for the USA during 1971-2010.

We suggest to consider this trend-breach as an effect of globalization after the end of the Cold War; that is, the demise of the Soviet Union, the unification of Germany, and the opening of China since the early 1990s (Leydesdorff & Sun, 2009). Note that the Triple Helix thesis was formulated in 1995 (Etzkowitz & Leydesdorff, 1995) as a reflection upon mainly the period before this global change.

Let us now focus on China. The increase of publications with an address in this country has been a major factor in the transformations at the global level during the second half of the 1990s and the 2000s.

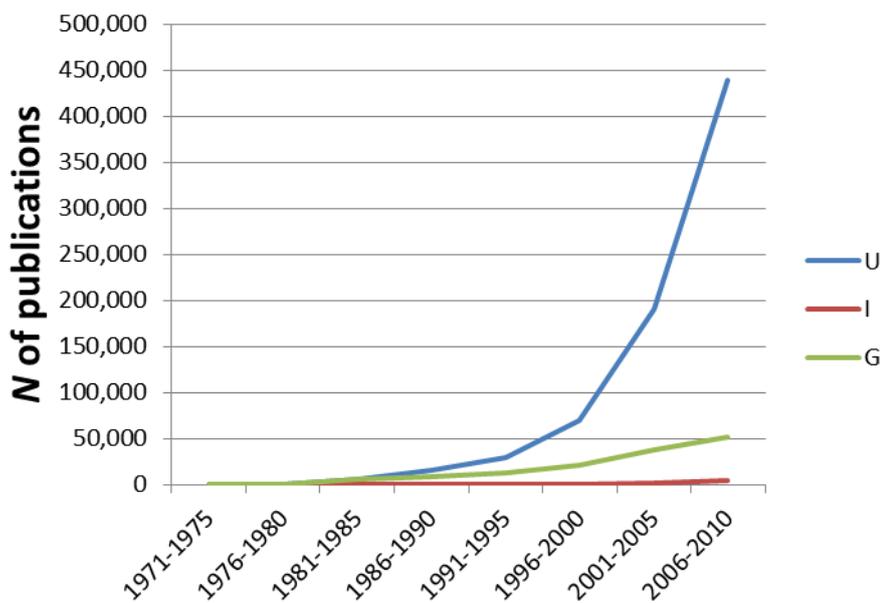

**Fig. 9**: Publication output of China in terms of U, I, and G; 1971-2010.

Figure 9 shows the well-known exponential growth curve of Chinese science (e.g., Jin & Rousseau, 2004; Moed, 2002; Plume, 2011; Zhou & Leydesdorff, 2006). After a lag-phase during the 1990s, the curve of academic publications entered into the log-phase during the first decade of the 2000s. We used here deliberately the data with CAS as G in order to show that the contributions of CAS(S) are not causing this change (as part of G). However, the attribution of CAS(S) addresses to G or U makes a difference for the Triple-Helix indicator (Figure 10).

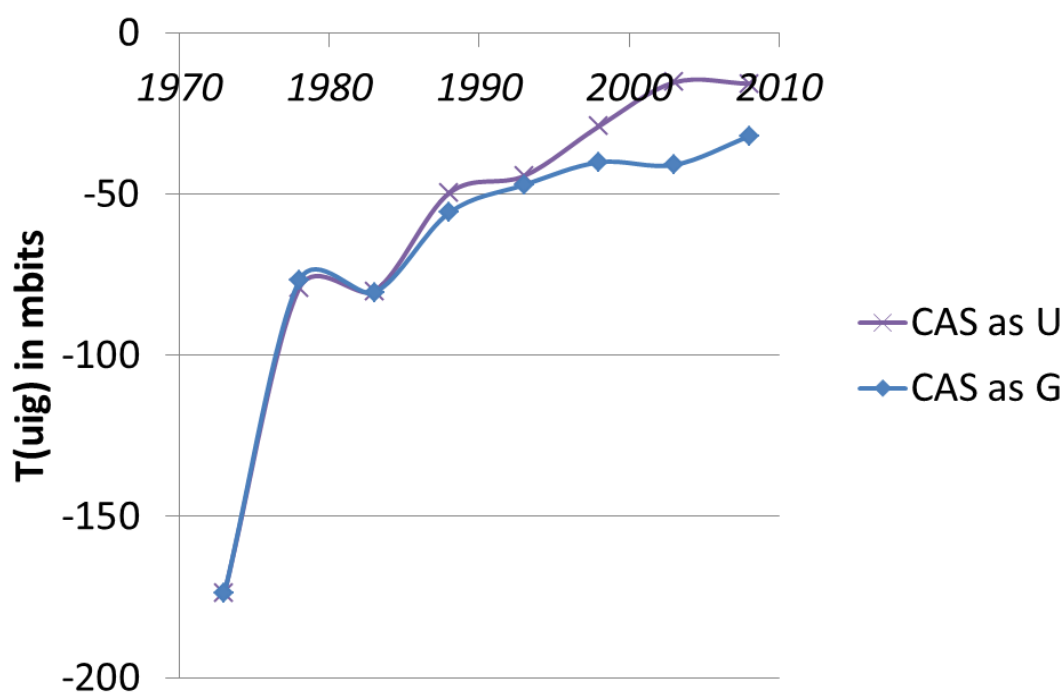

**Fig. 10**: The mutual information in UIG relations in China (CAS as U or G) during 1971-2010.

Figure 10 shows that China had a very strongly integrated system in terms of cross-sectional co-authorship relations during the 1970s. However, one should note that the numbers in WoS were low during this period (< 100). During the 1980s, the numbers were much larger and China was comparable with Indonesia nowadays in terms of the Triple-Helix indicator. During the 1990s, the attribution of "CAS as U" or "CAS as G" began to make a difference for triple-helix relations in China. This development was further reinforced during the 2000s.

The redefinition of CAS as part of the academic publication system has since been enhanced. For example, in "nano-technology", CAS can be considered as a leading research institute worldwide (Leydesdorff, in press). Figure 9 above already showed that "CAS as G" remains relatively small compared to the total volume of university publications in China.

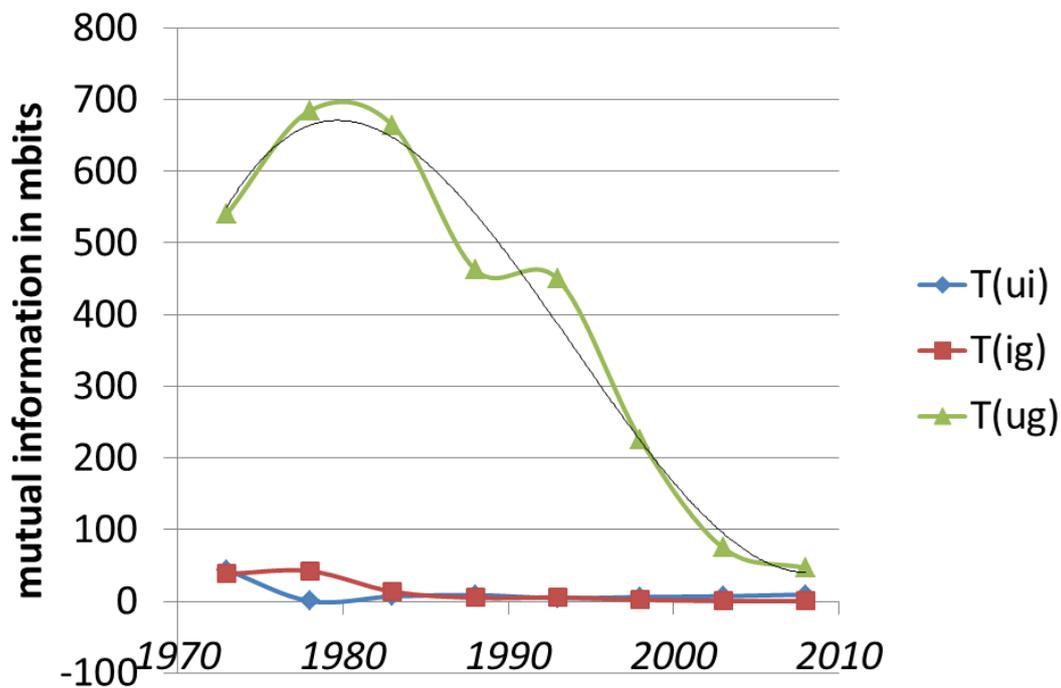

**Fig. 11**: Mutual information in bilateral relations when considering CAS as U for China, 1971-2010.

Further decomposition of the mutual information in bilateral relations in Figure 11 shows that U-G relations have continuously declined since the period 1980-1985 if CAS is considered as part of the university system. This uncoupling has eroded national integration of Triple-Helix relations even faster than in the USA. The industry participation in the publication system is still neglegible in China, and thus task differentiation in the various institutional domains has become prevalent. Academic publishing has boomed and international collaborations have become far more important than national and cross-sectoral ones (Leydesdorff & Jin, 2005).

**Discussion**

The Triple-Helix measure of synergy is a systems measure that is based on relational (in this case, coauthorship) data, but adds importantly to the analysis of co-authorship relations from another perspective. Relations among two or more parties are local, whereas a systems measure informs us about the state of the network that is shaped as the sumtotal of local relations. Different from other global measures (for example, betweenness centrality), the mutual information in three dimensions takes next-order loops in the data into account (Krippendorff, 2009). These next-order loops can generate redundancy. In the Triple Helix model this redundancy is considered as generated by the exchanges of meaning (on top of the information exchange) at the level of an "overlay" or hyper-cycle of communication (Etzkowitz & Leydesdorff, 2005; Leydesdorff, 2011).

In other words, a Triple Helix can generate a communication field as a second-order domain that remains dependent on the variation from which it is generated as the first-order domain, but the next-order field feeds potentially back on the variation in the first-order one because of its increasing systemness (Ivanova & Leydesdorff, in preparation). The effect of this systemness—its footprint—is indicated as potential reduction of the uncertainty that prevails.

One should note that U-I-G relations or, more generalized, knowledge-based innovation systems can be considered as complex systems with nonlinear feedback loops, in which many factors interact in a complex way with one another so that dynamic changes cannot be effectively forecasted. Therefore, the further extension with other domains of the Triple Helix (Lowe, 1982), quadruple or higher-order helix models (Carayannis & Campbell, 2009 and 2010), or even n-tuple helices (Leydesdorff, 2012) may also be valuable extensions for future research. Other indicators than coauthorship relations should also be used. Coauthorship relations were used in this study as a proxy for knowledge-based developments.

**Conclusion**

We found a different pattern for the various countries, and focused on the USA as a developed nation and China as a developing one. However, a common factor is globalization of the university publication system during the 1990s and 2000s which has loosened interactions with industry and government in terms of coauthorship relations.

Halffman & Leydesdorff (2010) noted that institutional incentives for universities have tended to become isomorphic in favor of publishing. This was further reinforced by the development of university ranking systems during the 2000s, such as the Academic Ranking of World Universities (ARWU) of the Shanghai Jiao Tong University in 2004, the THES-QS ranking in 2005, the Leiden rankings in 2008, and the rankings of Taiwan Higher Education and Accreditation Council also in 2008. For example, these rankings have not taken university patenting into account (Leydesdorff & Meyer, 2010).

At the country level, the interaction of university-industry-government can be measured in terms of the relations of mutual information, joint entropy, and configurational information. There are differences in the structure of Triple Helix relations between university, industry, and governments in developed and developing countries. The dynamic analysis of the Triple Helix indicator showed that the configurations among the three subsystems U-I-G became less negative over time in both developed and some developing nations. The relocation of CAS(S) in China from the public sector to the academic sector further enhances this development.

The Triple Helix model provides us with a point of reference for the analysis of the knowledge-based economy and scientific and technical innovation systems; U-I-G interactions provide valuable information for understanding innovation systems aggregated at the national level. For example, Figure 6 suggests that countries like India and Indonesia could profit from globalizing their innovation systems by loosening local UIG relations. Different countries appear to entertain different U-I-G patterns, and quantitative studies enable us to reveal patterns in the dynamics in this evolution.

## Acknowledgements

The first author is grateful to the financial support from NSFC Grants No. 71173187 and the discussion with Dr. Jiang Li and Mr. Star X. Zhao.